\documentclass[aps,prb,reprint,superscriptaddress,showpacs]{revtex4-1}
\usepackage{amsmath,amssymb}
\usepackage{amsbsy}
\usepackage{graphicx}

\newcommand{\vecA}{\mathbf{A}}
\newcommand{\vecd}{\mathbf{d}}
\newcommand{\abs}[1]{\left| #1 \right|}
\newcommand{\tildeVL}{\tilde{\mathbf{V}}_L}
\newcommand{\psiL}{(\tildeVL \nabla') \psi_0}
\newcommand{\vecjd}{\mathbf{j}'_d}
\newcommand{\etaxx}{\eta_x}
\newcommand{\etayy}{\eta_y}
\newcommand{\Phixnull}{\Phi_x^{(0)}}
\newcommand{\Phixone}{\Phi_x^{(1)}}
\newcommand{\fwithu}{f^2 \left( \frac{\rho}{\sqrt{u}} \right)}
\newcommand{\exponent}{\exp(-u\abs{x-x'} / \sqrt{s})}
\newcommand{\const}{\mathrm{const}}
\newcommand{\inftyint}{\int_{-\infty}^{+\infty}}

\newcommand{\lnsu}{\ln \frac{\sqrt{s}}{u}}
\newcommand{\Min}[1]{\min\limits_{#1}}
\newcommand{\vecrho}{\boldsymbol{\rho}}

\begin{document}
\renewcommand{\appendixname}{APPENDIX}

\title{Mismatch of conductivity
anisotropy in the mixed and normal states of type--II
superconductors.}
\author{A. A. Bespalov}
\affiliation{Institute for Physics of Microstructures,
 Russian Academy of Sciences, GSP-105, 603950, Nizhny Novgorod, Russia}
\affiliation{Univ. Bordeaux, LOMA, UMR 5798, F-33600 Talence, France}
\author{A. S. Mel'nikov}
\affiliation{Institute for Physics of Microstructures,
 Russian Academy of Sciences, GSP-105, 603950, Nizhny Novgorod, Russia}
\begin{abstract}
We have calculated the Bardeen-Stephen contribution to the vortex
viscosity for uniaxial anisotropic superconductors within the
time-dependent Ginzburg-Landau (TDGL) theory. We focus our
attention on superconductors with a mismatch of anisotropy of
normal and superconducting characteristics. Exact asymptotics for
the Bardeen-Stephen contribution have been derived in two
limits: (i) $l_{Eab} \ll \xi_{ab}$,  $l_{Ec} \ll \xi_{c}$ and (ii)
$l_{Ec} \gg \xi_{c}$, $l_{Eab} \lesssim \xi_{ab}$, where
$l_{Eab}$, $l_{Ec}$ and $\xi_{ab}$, $\xi_c$ are the electric field
penetration lengths and the coherence lengths in the $ab$ plane
and in the direction of the $c$ axis. Also we suggest a
variational procedure which allows us to calculate the vortex
 viscosity for superconductors with arbitrary parameters $\xi$ and $l_E$.
 The approximate analytical result is compared with numerical calculations.
 Finally, using a generalized TDGL theory, we prove that the viscosity
 anisotropy and, thus, the flux-flow conductivity anisotropy may depend on temperature.
\end{abstract}


\maketitle

\section{Introduction}

The existence of a non-zero electrical resistivity in type--II
superconductors in the mixed state is connected with the motion of
magnetic flux vortices.
 It can be observed in the presence of a sufficiently large transport current
  so that pinning is suppressed. In the stationary flux-flow regime the Lorentz
  force acting on an isolated vortex is balanced by the intrinsic viscous drag force:
%
%
\begin{equation}
    \frac{\phi_0}{c} \left( \mathbf{j}_{\mathrm{tr}} \times \mathbf{n} \right) = \eta \mathbf{V}_L
    \label{eq:vortex_motion}
\end{equation}
Here $\phi_0$ is the flux quantum, $\mathbf{j}_{\mathrm{tr}}$ is the
transport current density, $\mathbf{n}$ is the unit vector along the magnetic field,
$\mathbf{V}_L$ is the vortex velocity and $\eta$ is a viscosity coefficient.
As vortices move, the magnetic field in the sample becomes
nonstationary and a macroscopic electrical field $\mathbf{E}$ is
induced, which is connected with the transport current via Ohm's law:
 $\mathbf{E}=\mathbf{j}_{\mathrm{tr}}/\sigma$.
 For weak average magnetic fields,
 $B \ll H_{c2}$, where $H_{c2}$ is the upper critical field,
 the flux-flow conductivity $\sigma$ is
\begin{equation}
     \sigma = \frac{c^2 \eta}{B \phi_0}.
     \label{eq:sigma_iso}
\end{equation}
The presence of a finite conductivity implies
 that current flow is accompanied by dissipation.
 It has been shown that there are two main mechanisms
 of dissipation: losses due to relaxation of the order
  parameter\cite{Tinkham64} and ohmic losses associated with normal currents flowing through the vortex core.\cite{Bardeen+65}

For an anisotropic superconductor Eq. \eqref{eq:sigma_iso} is generalized as follows:
\begin{equation}
     \hat{\sigma} = \frac{c^2}{B \phi_0} \left(
    \begin{array}{rr}
        \eta_{yy} & -\eta_{yx} \\
        -\eta_{xy} & \eta_{xx}
    \end{array}
    \right)
    \label{eq:entangle}
\end{equation}
with the $z$-axis along the magnetic field. The
peculiar structure of the conductivity tensor is explained
 by the fact that the $x$-component of the electric field
 depends on the $y$-component of the vortex velocity, and vice versa.

It can be seen from Eq. \eqref{eq:entangle} that the flux-flow
conductivity is determined by the magnetic field and the viscosity
tensor $\hat{\eta}$. A rigorous approach to the problem of viscosity
 evaluation has been first suggested by Schmid\cite{Schmid66} and was later 
 developed by Gor'kov and Kopnin\cite{Gor'kov+71}
 (see also Ref. \onlinecite{Gor'kov+75} for review). Their method is based
 on the time-dependent Ginzburg-Landau (TDGL) theory. Within this model the
 flux-flow conductivity has been evaluated for isotropic superconductors
  in several papers.\cite{Hu+72, Hu72, Kupriyanov+72}. Both viscosity components
  due to order parameter relaxation  and ohmic losses
   (frequently called the Bardeen-Stephen contribution) have been derived.

Theoretical studies of free flux flow in anisotropic materials
have been stimulated by the discovery of
 high-temperature superconductors which appeared to possess rather
 strong anisotropy. A number of papers have addressed this problem
 using different models and approximations. \cite{Genkin+89,Ivlev+91,Hao+Clem91}
The procedure of viscosity calculation can be essentially
simplified in the limit of dirty uniaxial superconductors with the
ratio $s_0 = m_c \sigma_c/m_{ab} \sigma_{ab}$
 equal to unity. Here $\sigma_c$, $\sigma_{ab}$ and $m_c$, $m_{ab}$ are the normal conductivities
 and Cooper pair masses in the direction of the anisotropy axis $c$ and in the perpendicular
 $ab$-plane, respectively.
The condition $s_0=1$ allows to reduce the problem of anisotropic
vortex dynamics to an isotropic one by means of a scaling
transformation.\cite{Ivlev+91} Yet this is not true in the case
  $s_0 \neq 1$, i.e, for a mismatch of anisotropies of Cooper pair masses and normal conductivities.
   Such a mismatch is theoretically possible in  the relatively clean limit\cite{Kopnin_book} and it may have been
    experimentally observed in a new class of Fe-based pnictide superconductors.
     According to Ref. \onlinecite{Shirai+2010}, the ration $\sigma_{ab}/\sigma_c$ in
      $\mathrm{PrFeAsO_{0.7}}$ is close to 120, whereas $m_c/m_{ab}$ in the same compound
       is about 25, as determined in Ref. \onlinecite{Shirage+2009} from upper critical
       field measurements. In Refs. \onlinecite{Ni+2008,Wang+2009} anisotropies of the
       same order in $\mathrm{Ba_{1-x}K_x Fe_2 As_2}$ are reported. However, existing experimental
        data for the pnictides are contradictory. In Ref. \onlinecite{Tanatar+2009} a relatively
        low resistivity anisotropy in $\mathrm{BaFe_2As_2}$ is given: $\sigma_{ab}/\sigma_{c} \sim 2-3$.
        In some works\cite{Shirage+2009,Tanatar+2009BaFe2As2} an anisotropy mismatch has not been clearly detected.
Previous calculations of the vortex viscosity tensor accounted for
the anisotropy mismatch only on the basis of a simplified model of
a step-like order parameter profile within the vortex core
.\cite{Genkin+89} Of course, a detailed comparison with
experimental data demands these calculations to be generalized for
a more realistic order parameter profile.

In this paper we evaluate analytically the
 viscosity tensor for a realistic gap profile within the core, focusing
our attention on the case $s_0 \neq 1$ and considering both
standard TDGL model and its generalization for superconductors
with a finite gap.\cite{Watts-Tobin+81,Kopnin_book} In section
\ref{sec:maineq} we derive the basic equations following the
approach of Gor'kov and Kopnin.\cite{Gor'kov+75} In Section
\ref{sec:approximate} we develop approximate methods based on
different assumptions about the ratio of the electric field
penetration depth to the coherence length. The results of
preceding works\cite{Hu+72, Hu72, Genkin+89} are
revised and improved. In the end of this section we consider a
variational principle which provides us with a simple general
relation for the Bardeen--Stephen contribution. In section
\ref{sec:GTDGL} the problem is considered in the framework of
a generalized TDGL theory. Here we derive our main
result: we predict that the flux-flow conductivity anisotropy may
depend on temperature in superconductors with the parameter $s_0
\neq 1$.

\section{Basic equations}
\label{sec:maineq}

Following Gor'kov and Kopnin,\cite{Gor'kov+75} we start the
analysis of vortex motion with the TDGL equation for the
superconducting order parameter $\psi$:
\begin{equation}
    \gamma \left( \hbar \frac{\partial \psi}{\partial t} + 2ie \Phi \psi \right) = -\frac{ \delta F}{\delta \psi^*},
    \label{eq:TDGL}
\end{equation}
\begin{eqnarray*}
   F= \int{\left[ \left(i\hbar \nabla - \frac{2e}{c} \mathbf{A} \right) \psi^* \frac{\hat{m}^{-1}}{2} \left(-i\hbar \nabla - \frac{2e}{c} \mathbf{A} \right)\psi \right.} & & \\
  \qquad \left. {}+ a \abs{\psi}^2 + \frac{1}{2} b \abs{\psi}^4 \right] d^3 \mathbf{r}. & & 
\end{eqnarray*}
Here $F$ is the usual GL free energy, $\gamma$ is a relaxation
constant and $\mathbf{A}$ and $\Phi$ are the vector and scalar
potentials, respectively. We consider uniaxial
 anisotropic superconductors, so the Cooper pair mass $\hat{m}$
 is a tensor with components $m_{ij} = m_{ab} (\delta_{ij} + \mu \nu_i \nu_j)$
  where $\boldsymbol{\nu}$ is the unit vector along the $c$-axis, $\mu = m_c/m_{ab} - 1$.
  Eq. \eqref{eq:TDGL} is supplemented by the equation for the current density
\begin{equation}
    \mathrm{div} \, \mathbf{j} = 0,
    \label{eq:divj}
\end{equation}
where
\begin{equation}
    \mathbf{j} = 2e \abs{\psi}^2 \hat{m}^{-1} \left( \hbar \nabla \theta - \frac{2e}{c} \vecA \right) - \hat{\sigma}_n \left( \nabla \Phi + \frac{1}{c} \frac{\partial \vecA}{\partial t}
    \right).
    \label{eq:j}
\end{equation}
Here $\theta = \mathrm{arg}(\psi)$ and $\hat{\sigma}_n$ is the
normal-state conductivity tensor with components
$\sigma_{nij}=\sigma_{ab}\delta_{ij} + (\sigma_c - \sigma_{ab})
\nu_i \nu_j$. For simplicity we will consider only superconductors
with a large Ginzburg-Landau parameter $\kappa = \lambda/\xi \gg
1$, where $\lambda$ is the London penetration length and $\xi$ is
the superconducting coherence length. One can prove that
$\abs{2e \vecA/c}/\abs{\hbar \nabla \theta} \ll 1$
 at distances much smaller than $\lambda$ from the vortex axis
in the gauge where $\vecA = 0$ on the vortex axis and $\mathrm{div} \vecA = 0$.
Imposing the additional condition $l_E^2/\lambda \xi \ll~1$, where
$l_E$ is the electric field penetration depth (see Eq.
\eqref{eq:xi_lE}), one can neglect the term
\[ \frac{1}{c} \frac{\partial \vecA}{\partial t} \]
in Eq. \eqref{eq:j}.

\begin{figure}[t]
  \hspace{-4.5cm}
  \includegraphics{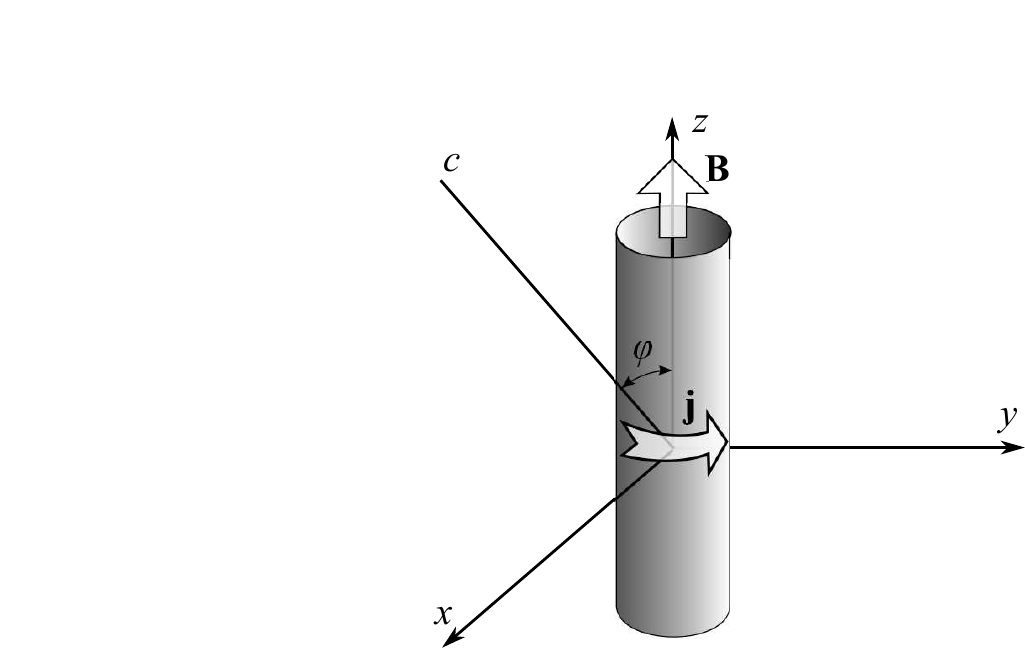}
  \caption{The coordinate frame}
  \label{fig:frame}
\end{figure}

Let us consider the orientation of the internal magnetic field at an angle $\varphi$ to the crystallographic $c$-axis. We choose the coordinate frame $(x,\, y,\, z)$ with the $z$-axis coinciding with the vortex axis and with the $c$-axis lying in the $xz$-plane (see Fig. \ref{fig:frame}). In this frame the functions $\psi$ and $\Phi$ do not depent on $z$, and the tensor $\hat{\eta}$ is diagonal.

The derivation of the force balance equation \eqref{eq:vortex_motion} and the explicit expression for the viscosity may be found in Refs. \onlinecite{Gor'kov+75,Genkin+89}. However, in Appendix \ref{app:basic} we outline the calculations for the reader's convenience.

The components of the Bardeen-Stephen contribution, $\hat{\eta}_{\mathrm{oh}}$, are given by
\begin{equation}
    \etaxx = -2 \frac{ \left| a \right|}{b} \gamma \hbar \int{ f^2(\rho_1) \frac{y_1}{\rho_1^2} \left( u^2 \Phi_x - \frac{y_1}{\rho_1^2} \right) \, dx_1 dy_1 },
    \label{eq:eta_xx}
\end{equation}
\begin{equation}
    \etayy = -2 \frac{ \left| a \right|}{b} \gamma \hbar \int{ f^2(\rho_1) \frac{x_1}{\rho_1^2} \left( u^2 \Phi_y - \frac{x_1}{\rho_1^2} \right) \, dx_1 dy_1 }.
    \label{eq:eta_yy}
\end{equation}
Here
\begin{eqnarray}
	& \eta_x = [m(\varphi)/m_{ab}]^{1/2} (\eta_{\mathrm{oh}})_{xx} &	\label{eq:etax_def}   \\
  & \eta_y = [m_{ab}/m(\varphi)]^{1/2} (\eta_{\mathrm{oh}})_{yy}, & \label{eq:etay_def},
\end{eqnarray}
\[ m(\varphi) = \frac{m_{ab} (1+\mu)}{1 + \mu \cos^2 \varphi}, \qquad u = \xi_{ab}/l_{Eab}, \]
\begin{equation}
	 (x_1,y_1) = \xi_{ab}^{-1} \left( \sqrt{\frac{m(\varphi)}{m_{ab}}} x, y \right), \qquad \rho_1 = \sqrt{x_1^2 + y_1^2},
	 \label{eq:x1y1}
\end{equation}
$\xi_{ab}$ and $l_{Eab}$ are the coherence length and the electric field penetration depth in the $ab$-plane, respectively:
\begin{equation}
	\xi_{ab} = \sqrt{\frac{\hbar^2}{2 m_{ab} \abs{a}}}, \qquad
     l_{Eab}=\left[ \hbar \sigma_{ab}/\left( 8 e^2 \gamma \frac{\abs{a}}{b}\right) \right]^{1/2}.
     \label{eq:xi_lE}
\end{equation}
The function $f(\rho)$ describes the profile of the dimensionless order parameter modulus in a static isotropic vortex. This function satisfies the relation
\begin{equation}
    \frac{1}{\rho} \frac{d}{d\rho} \left( \rho \frac{df}{d \rho} \right) - \frac{f}{\rho^2} + f - f^3 = 0,
    \label{eq:f}
\end{equation}
which follows from Eqs. \eqref{eq:psi0} and \eqref{eq:f_def}. The boundary conditions are $f(0)=0$, $f(\infty) = 1$. The functions $\Phi_x$ and $\Phi_y$ in Eqs. \eqref{eq:eta_xx} and \eqref{eq:eta_yy} should be determined from the linear equations
\begin{equation}
    s \frac{\partial^2 \Phi_x}{\partial x_1^2} + \frac{\partial^2 \Phi_x}{\partial y_1^2} = \left( u^2 \Phi_x - \frac{y_1}{\rho_1^2} \right) f^2(\rho_1),
    \label{eq:Phi_x}
\end{equation}
\begin{equation}
    s \frac{\partial^2 \Phi_y}{\partial x_1^2} + \frac{\partial^2 \Phi_y}{\partial y_1^2} = \left( u^2 \Phi_y - \frac{x_1}{\rho_1^2} \right) f^2(\rho_1),
    \label{eq:Phi_y}
\end{equation}
where
\begin{equation}
    s(\varphi) = 1 + \left( \frac{m_c \sigma_c}{m_{ab} \sigma_{ab}} - 1 \right) \frac{\sin^2{\varphi}}{1 + \mu \cos^2{\varphi}} >0.
    \label{eq:s}
\end{equation}
The electric potential can be expressed in terms of $\Phi_x$ and $\Phi_y$ via
\begin{equation}
	 \Phi = \left( \Phi_x \sqrt{\frac{m(\varphi)}{m_{ab}}} V_{Lx} - \Phi_y V_{Ly}  \right) \frac{4 \gamma e \hbar}{b \sigma_{ab}} \sqrt{\frac{\left| a \right|}{2 m_{ab}}}. 
	 \label{eq:PhixPhiy}
\end{equation}

Note that there is a relation connecting the components $\eta_x$ and $\eta_y$:
\begin{equation}
    \etayy(s,u) = \etaxx\left( \frac{1}{s},\frac{u}{\sqrt{s}} \right).
    \label{eq:connection}
\end{equation}
In the next section we consider some limiting cases.
\section{Approximate solutions}
\label{sec:approximate}
\subsection{The $l_E \ll \xi$ limit}
\label{sub:large_u}

Consider such materials that the electric field penetration length is much smaller than the coherence length:
\begin{equation}
    l_{Eab} \ll \xi_{ab}, \quad l_{Ec} \ll \xi_{c}.
    \label{eq:small_lE}
\end{equation}
This limiting case is more close to gapless superconductors with a high concentration of magnetic impurities which are characterized by the ration $\xi/l_E = \sqrt{12}$. The conditions \eqref{eq:small_lE} impose the following restrictions on the parameters $s$ and $u$: $u \gg 1$, $s \ll u^2$. In this subsection we will analyse the case $s \lesssim 1$. The case $1 \ll s \ll u^2$ can be considered in a similar way by dividing Eqs. \eqref{eq:Phi_x} and \eqref{eq:Phi_y} by $s$. We shall search the asymptotics of the viscosity when $u \to \infty$ neglecting small terms of order higher than $u^{-2}$ (however, it will be shown that one should keep terms of the order of $u^{-2}$).

Our approximation is based on the fact that the characteristic
length scale for the functions $\Phi_x$ and $\Phi_y$ is $u^{-1}$.
Hence, the unknown functions reach their asymptotic behavior at
distances $\rho \ll 1$ from the vortex axis, where the order
parameter profile $f(\rho)$ is well approximated by the first
several terms of its Taylor series:
\[ f^2(\rho) \approx k_2 \rho^2 + k_4 \rho^4 + k_6 \rho^6. \]
We substitute this expansion into Eq. \eqref{eq:Phi_x} and introduce the new
variables $\tilde{\boldsymbol{\rho}} = \boldsymbol{\rho}_1
\sqrt{u}$, $\tilde{\Phi}_x = \Phi_x u^{3/2}$:
\begin{eqnarray}
    \lefteqn{\frac{\partial^2 \tilde{\Phi}_x}{\partial \tilde{y}^2} + s \frac{\partial^2 \tilde{\Phi}_x}{\partial \tilde{x}^2}} & & \nonumber \\
    & & =\left( k_2 \tilde{\rho}^2 + k_4 \frac{\tilde{\rho}^4}{u} + ...  \right) \left( \tilde{\Phi}_x - \frac{\tilde{y}}{\tilde{\rho}^2} \right).
    \label{eq:tildePhi_x}
\end{eqnarray}
Further the tilde over $\tilde{x}$ and $\tilde{y}$ will be omitted. The solution of Eq. \eqref{eq:tildePhi_x} can be expanded in the powers of $u^{-1}$:
\begin{equation}
    \tilde{\Phi}_x = \Phixnull + u^{-1} \Phixone + R_x,
    \label{eq:Phix_expand1}
\end{equation}
where $\Phixnull$ and $\Phixone$ satisfy the following relations:
\begin{equation}
    \frac{\partial^2 \Phixnull}{\partial y^2} + s \frac{\partial^2 \Phixnull}{\partial x^2} = k_2 \rho^2 \Phixnull - k_2 y,
    \label{eq:Phixnull}
\end{equation}
\begin{equation}
    \frac{\partial^2 \Phixone}{\partial y^2} + s \frac{\partial^2 \Phixone}{\partial x^2} = k_2 \rho^2 \Phixone + k_4 \rho^4 \left( \Phixnull - \frac{y}{\rho^2} \right),
    \label{eq:Phixone}
\end{equation}
and $R_x$ is a remainder term. It is proved in Appendix \ref{app:largeu} that an analogous expansion can be made in the integral in the rhs of Eq. \eqref{eq:eta_xx}:
\begin{equation}
    \eta_{x} = -2 \frac{\abs{a}}{b} \gamma \hbar \left[ \frac{I_{1x}(s)}{u}   + \frac{I_{2x}(s)}{u^2} + o(u^{-2}) \right],
    \label{eq:etaxx_final}
\end{equation} 
where
\begin{equation}
	I_{1x}(s) = \int k_2 y \left( \Phixnull - \frac{y}{\rho^2} \right) dx \, dy,
	\label{eq:I_1x_def}
\end{equation}
\begin{equation}
	I_{2x}(s) = \int \!\! \frac{y}{\rho^2} \! \left[ k_4 \rho^4 \left( \Phixnull - \frac{y}{\rho^2} \right) + k_2 \rho^2 \Phixone \right] dx \,dy.
	\label{eq:I_2x_def}
\end{equation}
The viscosity component $\etayy$ can be calculated similarly:
\begin{equation}
    \eta_{y} = -2 \frac{\abs{a}}{b} \gamma \hbar \left[ \frac{I_{1y}(s)}{u}   + \frac{I_{2y}(s)}{u^2} + o(u^{-2}) \right].
    \label{eq:etayy_final}
\end{equation}
Using Eq. \eqref{eq:connection} we obtain
\begin{equation}
	 I_{1y}(s) = I_{1x}(s^{-1}) \sqrt{s}, \qquad I_{2y}(s) = I_{2x}(s^{-1}) s.
	 \label{eq:I_xy}
\end{equation}
In principle, the functions $I_{1x}(s)$ and $I_{2x}(s)$ can be determined by numerical calculations, however, in section \ref{sub:variational} we present some analytical expressions for these functions.

In Ref. \onlinecite{Genkin+89} the $u \gg 1$ limit was considered using the Bardeen-Stephen model.\cite{Bardeen+65} This approach is essentially based on the assumption about a step-like order parameter profile within the core and does not allow to obtain a leading term of the order of $u^{-1}$ in the expansion \eqref{eq:etaxx_final}.

The particular case $s=1$ has been considered in a number of works mentioned above.\cite{Schmid66,Hu+72,Hu72,Kupriyanov+72} It corresponds to isotropic superconductors, or anisotropic superconductors with no anisotropy mismatch: $(m_c \sigma_c)/(m_{ab} \sigma_{ab}) = 1$. If $s=1$, Eqs. \eqref{eq:Phixnull} and \eqref{eq:Phixone} can be solved exactly:
\[ \Phixnull = \frac{1-\exp( -\sqrt{k_2} \rho^2 / 2)}{\rho^2} y, \]
\[ \Phixone =  \frac{k_4 y}{u k_2} \left( \frac{1}{4} + \frac{\sqrt{k_2} \rho^2}{8} \right) \exp( -\sqrt{k_2} \rho^2 / 2). \]
After some integration we obtain a simple relation for the viscosity:
\begin{equation}
     \etaxx = \etayy = 2 \pi \frac{ \left| a \right|}{b} \gamma \hbar \alpha_2(u),
    \label{eq:eta_short}
\end{equation}
\begin{equation}
    \alpha_2(u) \approx \frac{\sqrt{k_2}}{u} + \frac{k_4}{2 k_2 u^2} = \frac{0.583}{u} - \frac{1}{8 u^2}.
    \label{eq:alpha_2}
\end{equation}
Here the value $\sqrt{k_2} = 0.583$ was taken from Ref. \onlinecite{Hu72}, and the relation $k_4 = -k_2/4$ follows from Eq. \eqref{eq:f}.

It is appropriate to recall here the result obtained by Hu:\cite{Hu72}
\begin{equation}
    \alpha_2 = \frac{\mathrm{K}_0(\delta u)}{\delta u \cdot \mathrm{K}_1(\delta u)},
    \label{eq:alpha_2_approx}
\end{equation}
where $\mathrm{K}_0$ and $\mathrm{K}_1$ are the modified Bessel functions of an imaginary argument and $\delta$ is a fitting parameter. Eq. \eqref{eq:alpha_2_approx} was derived from the exact solution of Eq. \eqref{eq:Phi_initial} with an approximate order parameter profile:
\begin{equation}
    f(\rho) = \frac{\rho}{\sqrt{\delta^2 + \rho^2}}.
    \label{eq:f_approx}
\end{equation}
According to Schmid\cite{Schmid66} and Hu, the optimal value of $\delta$ is $\sqrt{2}$ which follows from a variational principle. We can compare different values of $\alpha_2(u)$. When $u = \sqrt{12}$
Eq. \eqref{eq:alpha_2_approx} yields $\alpha_2 = 0.186$, Eq. \eqref{eq:alpha_2} yields $\alpha_2 = 0.158$, while the numerical result is $\alpha_2 = 0.159$.\cite{Kupriyanov+72} Our formula gives an error less than 1\%. If we keep only the term of order $u^{-1}$ in Eq. \eqref{eq:alpha_2}, we will get a 6\% error which increases with decreasing $u$.
\subsection{The $l_{Ec} \gg \xi_c$ limit.}
\label{sub:bigs}

Consider the range of parameters $s \gg u^2$ and $u \gtrsim 1$. In terms of $l_{E}$, $\xi$ and $\varphi$ these conditions read
\[ l_{Ec} \gg \xi_{c}, \qquad l_{Eab} \lesssim \xi_{ab}, \qquad \cos^{2} \varphi \ll \frac{\sigma_c l_{Eab}^2}{\sigma_{ab} \xi_{ab}^2}. \]
Thus, the magnetic field must make a small angle with the $ab$-plane.

When $s \gg u^2$, the term $u^2 \Phi_x$ in Eq. \eqref{eq:Phi_x} is negligible compared to $y_1/\rho_1^2$ in the region $\rho \ll \sqrt{s}/u$, so we immediately obtain from Eq. \eqref{eq:eta_xx}
\[ \etaxx \sim \ln{s/u^2}. \]
More complicated calculations, which can be found in Appendix \ref{app:sggu2}, yield
\begin{equation}
    \etaxx = 2\pi \hbar \gamma \frac{\left| a \right|}{b} \left( \ln \frac{\sqrt{s}}{u} - 1.475 \right),
    \label{eq:eta_x_s_final}
\end{equation}
\begin{equation}
    \etayy = 2\pi \hbar \gamma \frac{\left| a \right|}{b} \left( \ln \frac{\sqrt{s}}{u} - 0.475 \right).
    \label{eq:eta_y_s_final}
\end{equation}
Note that in Ref. \onlinecite{Genkin+89} in the $u \ll 1$ limit
similar expressions containing $\ln u^{-1}$ have been derived.
This similarity is not accidental: the presence of the logarithm
$\ln (l_E/\xi)$ is a characteristic feature of the $l_E \gg \xi$
limit.
\subsection{A variational principle}
\label{sub:variational}

In this subsection we suggest a simple variational procedure for
the calculation of the viscosity tensor. According to Ref. 
\onlinecite{Gor'kov+75} a general expression for the dissipation
function $W[\Phi]$ reads:
\begin{equation}
    W[\Phi] = \nabla \Phi \hat{\sigma}_n \nabla \Phi + \frac{2 \gamma}{\hbar} \abs{ \hbar \frac{\partial \psi}{\partial t} + 2ie\Phi \psi}^2.
    \label{eq:dissipative}
\end{equation}
The electric potential should be found from Eq.
\eqref{eq:Phi_initial} which can be viewed as a condition of zero
variational derivative of the functional
\[ \int_{z=0} W[\Phi] d^2 \boldsymbol{\rho}. \]
Thus, the minimum of the functional above equals the loss power
per unit length of a moving vortex:
\begin{equation}
    \mathbf{V}_L \hat{\eta} \mathbf{V}_L = \min\limits_{\Phi} \int_{z=0} W[\Phi] d^2 \boldsymbol{\rho}.
    \label{eq:variational}
\end{equation}
This relation allows us to apply the direct variational method to our problem.

For the sake of convenience we rewrite Eq. \eqref{eq:variational}
in our rescaled coordinate frame separately for both components of
the Bardeen-Stephen contribution:
\begin{equation}
    \etaxx = \tilde{\eta}(s,1,u), \qquad \etayy = \tilde{\eta}(1,s,u),
    \label{eq:eta_var'}
\end{equation}
\begin{eqnarray}
    &\tilde{\eta}(s_x,s_y,u) = 2 \frac{ \left| a \right|}{b} \gamma \hbar u^2 \cdot \Min{\phi} \int \left[ s_x \left(\frac{\partial \phi}{\partial x} \right)^2 \right. & \nonumber \\
    &\left. + s_y \left(\frac{\partial \phi}{\partial y} \right)^2 + \frac{f^2(\rho)}{u^2} \left( u^2 \phi - \frac{y}{\rho^2} \right)^2 \right] dx \, dy.&
    \label{eq:tilde_eta}
\end{eqnarray}
Eqs. \eqref{eq:eta_var'} - \eqref{eq:tilde_eta} have two important consequences. First, the viscosity tensor is positively defined when $\sigma_{ab}>0$ and $\sigma_c>0$. Second, the components of $\hat{\eta}_{\mathrm{oh}}$ increase as the conductivity increases:
\[ \frac{\partial \eta_i}{\partial \sigma_{ab}}>0, \quad \frac{\partial \eta_i}{\partial \sigma_c}>0, \quad i=x,y. \]

We can obtain an upper estimate for the viscosity components if we
substitute a trial function into Eq. \eqref{eq:tilde_eta}. In
order to find an appropriate trial function consider the exact
equation for $\phi$:
\begin{equation}
    s_x \frac{\partial^2 \phi}{\partial x^2} + s_y \frac{\partial^2 \phi}{\partial y^2} = \left( u^2 \phi - \frac{y}{\rho^2} \right) f^2(\rho).
    \label{eq:phi}
\end{equation}
The solution of this equation is an even function of $x$ and an odd function of $y$, so its Fourier series has the form
\begin{equation}
    \phi = \sum_{n=0}^{\infty}{\phi_{2n+1}}(\rho)\sin (2n+1)\chi,
    \label{eq:series}
\end{equation}
where $\chi$ is the polar angle in the $xy$ plane. When $\rho$ is sufficiently large, $\phi \approx y/(u^2 \rho^2)$, that means, that the series in Eq. \eqref{eq:series} contains only the first term. Thus, the trial function
\begin{equation}
    \phi_t = \frac{4\tilde{\phi}(\rho)}{s_x+3s_y} \sin{\chi}
    \label{eq:trial}
\end{equation}
has the correct parity and the correct asymptotics. Let us substitute this function into Eq. \eqref{eq:tilde_eta}:
\begin{eqnarray}
    & & \tilde{\eta} \approx 2 \frac{ \left| a \right|}{b} \gamma \hbar \pi \tilde{u}^2 \Min{\tilde{\phi}} \int_0^{\infty} \rho \left[ \left(\frac{d \tilde{\phi}}{d \rho} \right)^2 + \frac{\tilde{\phi}^2}{\rho^2} \right. \nonumber \\
    & & \qquad \qquad \qquad \left. + \frac{f^2(\rho)}{\tilde{u}^2} \left(\tilde{u}^2 \tilde{\phi} - \frac{1}{\rho} \right)^2 \right] d\rho,
    \label{eq:tilde_eta_iso}
\end{eqnarray}
where
\[ \tilde{u} = u \left(\frac{s_x}{4}+\frac{3s_y}{4} \right)^{-1/2}. \]
The differential equation for $\tilde{\phi}$ is
\begin{equation}
     -\frac{1}{\rho} \frac{d}{d\rho} \left( \rho \frac{d\tilde{\phi}}{d\rho} \right) + \frac{\tilde{\phi}}{\rho^2} + f^2(\rho) \left( \tilde{u}^2\tilde{\phi} - \frac{1}{\rho} \right) = 0.
     \label{eq:tilde_phi}
\end{equation}
Note that we obtain exactly the same equation if we substitute $\Phi_x=\tilde{\phi}(\rho) \sin{\chi}$ into Eq. \eqref{eq:Phi_x} when $s=1$ and $u=\tilde{u}$. This means that the trial function \eqref{eq:trial} reduces our problem to an isotropic one. Unfortunately, an exact solution of Eq. \eqref{eq:tilde_phi} is unknown. However, Schmid \cite{Schmid66} found a solution with an approximate order parameter profile [see Eq. \eqref{eq:f_approx}]:
\[ \tilde{\phi} = \frac{\mathrm{K}_1(\tilde{u}\delta) \delta - \sqrt{\delta^2 + \rho^2} \mathrm{K}_1(\tilde{u} \sqrt{\delta^2 + \rho^2})}{\delta \mathrm{K}_1(\tilde{u} \delta) \tilde{u}^2 \rho}. \]
Using this function and the expression \eqref{eq:f_approx} for $f$ we can calculate the rhs of Eq.\eqref{eq:tilde_eta_iso}:
\begin{equation}
    \tilde{\eta} \approx 2 \frac{ \left| a \right|}{b} \gamma \hbar \pi \frac{\mathrm{K}_0(\delta \tilde{u})}{\delta \tilde{u}  \mathrm{K}_1(\delta \tilde{u})}.
    \label{eq:tilde_eta_final}
\end{equation}
We take $\delta=f'(0)^{-1}=k_2^{-1/2}$ in order to obtain the correct asymptotics when $u \to \infty$, $s=1$ (this asymptotics is determined by $f'(0)$, see subsection \ref{sub:large_u}). Finally, combining \eqref{eq:eta_var'} and \eqref{eq:tilde_eta_final} we get approximate relations for the components of $\hat{\eta}'$:
\begin{equation}
    \etaxx \approx 2 \pi \frac{\abs{a}}{b} \gamma \hbar \frac{f'(0)}{2u}\sqrt{s+3} \frac{\mathrm{K}_0 \left( \frac{2u}{f'(0) \sqrt{s+3}} \right)}{\mathrm{K}_1 \left( \frac{2u}{f'(0) \sqrt{s+3}} \right)},
    \label{etaxx_universal}
\end{equation}
\begin{equation}
    \etayy \approx 2 \pi \frac{\abs{a}}{b} \gamma \hbar \frac{f'(0)}{2u}\sqrt{3s+1} \frac{\mathrm{K}_0 \left( \frac{2u}{f'(0) \sqrt{3s+1}} \right)}{\mathrm{K}_1 \left( \frac{2u}{f'(0) \sqrt{3s+1}} \right)}.
    \label{etayy_universal}
\end{equation}
No restrictions on the parameters $s$ and $u$ are implied here.

Let us check if these relations are in accordance with the results from subsections \ref{sub:large_u} and \ref{sub:bigs}. Expanding $\eta_x$ in the form \eqref{etaxx_universal} in the powers of $u^{-1}$ when $u \gg 1$ and $s \lesssim 1$ we obtain the following expressions for the coefficients $I_{1x}$ and $I_{2x}$, which were introduced in subsection \ref{sub:large_u} (see Eq. \eqref{eq:etaxx_final}):
\begin{equation}
	I_{1x}(s) = - \frac{\pi \sqrt{k_2} \sqrt{s+3}}{2}, \qquad I_{2x}(s) = \frac{\pi k_2 (s+3)}{8}.
	\label{eq:Ix_var} 
\end{equation}
When $s=1$
\[ \etaxx=\etayy = 2 \pi \frac{\abs{a}}{b} \gamma \hbar \frac{f'(0)}{u} + O(u^{-2}), \]
which should be compared with Eq. \eqref{eq:alpha_2}. The perfect
agreement between the exact and approximate result is not
surprising, because the trial function \eqref{eq:trial} is the
exact solution of our variational problem in the isotropic case.

In order to check whether Eqs.
\eqref{etaxx_universal} and \eqref{etayy_universal} are applicable for $s \neq 1$
we used numerical calculations. We solved Eq. \eqref{eq:phi}
in the region $x>0$, $y>0$ with the boundary conditions
\[ \frac{\partial \phi}{\partial x} \biggl|_{x=0} = 0, \qquad \phi \biggl|_{y=0} = 0. \]
A sufficiently large $450\times 450$ mesh with a $0.03 \times 0.03$ unit cell has been used. The numerical algorithm applied was the method of steepest descent. After the determination of the function $\phi(\vecrho)$ numerical integration has been performed.

When $s=0$ Eqs. \eqref{eq:Ix_var} and \eqref{eq:I_xy} give
\[ \etaxx = 2 \frac{\abs{a}}{b} \gamma \hbar \frac{1.59}{u} + O(u^{-2}), \]
\[ \etayy = 2 \frac{\abs{a}}{b} \gamma \hbar \frac{0.92}{u} + O(u^{-2}). \]
These analytical expressions are in a good agreement  with the
asymptotics derived by numerical calculations:
\[ \etaxx \! = 2 \frac{\abs{a}}{b} \gamma \hbar \frac{1.58}{u} +O(u^{-2}), \]
\[ \etayy \! = 2 \frac{\abs{a}}{b} \gamma \hbar \frac{0.86}{u} + O(u^{-2}).\]
When $s \gg u^2$ Eqs. \eqref{etaxx_universal} and \eqref{etayy_universal} give
\[ \etaxx \approx \etayy \approx 2 \pi \hbar \gamma \frac{\left| a \right|}{b} \ln \frac{\sqrt{s}}{u}, \]
which coincides with the main logarithmic term in Eqs.
\eqref{eq:eta_x_s_final} and \eqref{eq:eta_y_s_final}.

One can see that the agreement between the exact and approximate asymptotics is quite well. This is a strong argument in favor of the applicability of Eqs. \eqref{etaxx_universal} and \eqref{etayy_universal} for intermediate values of $s$ and $u$.

\begin{figure}[t]
  \includegraphics[scale=0.3]{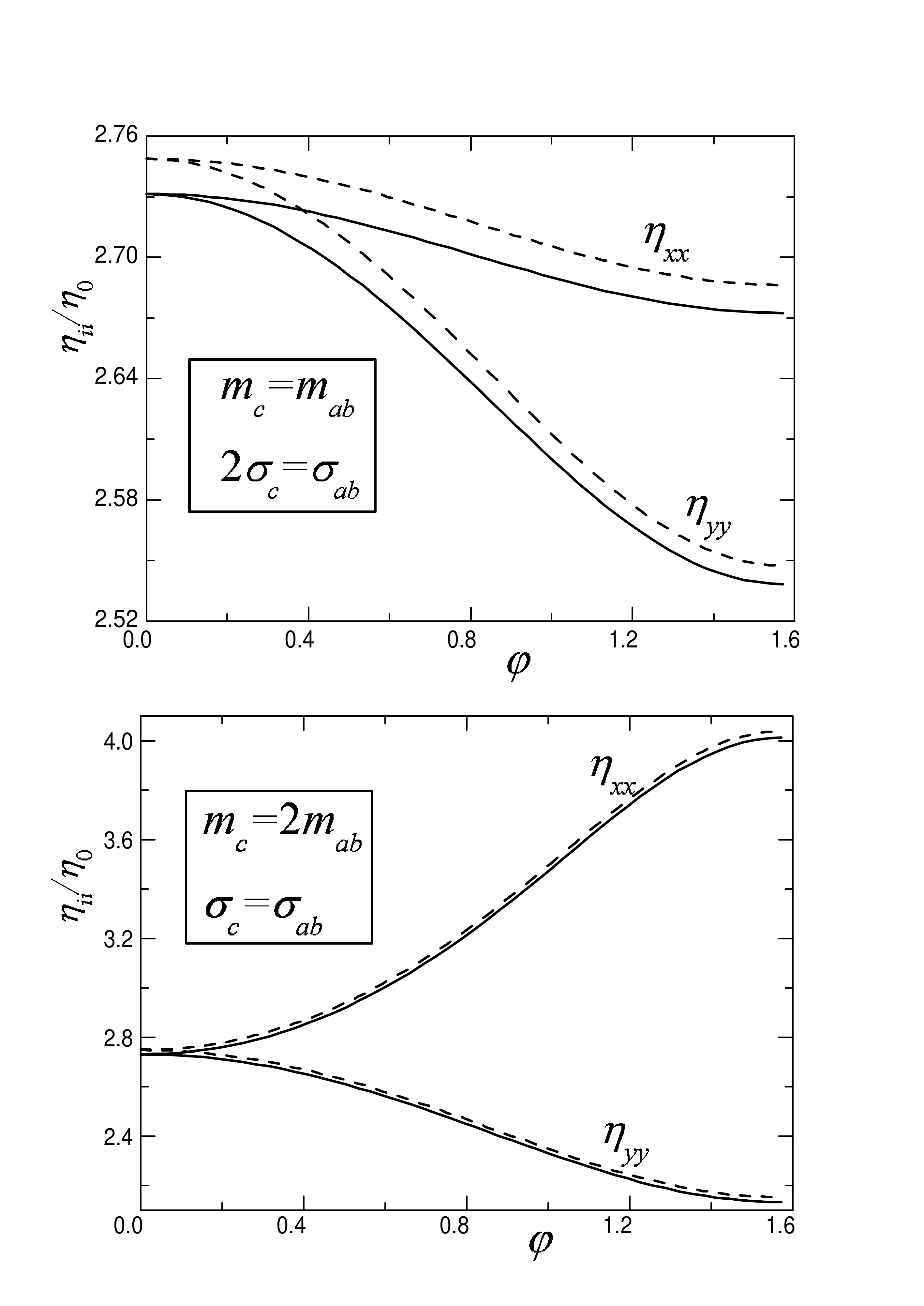}
  \caption{The $\varphi$ dependencies of the viscosity components. Solid lines correspond to analytical results (Eqs. \eqref{etaxx_universal} and \eqref{etayy_universal}), dashed lines show the results of numerical simulations. Here $\eta_{xx}$ and $\eta_{yy}$ are measured in the units $\eta_0=\hbar \gamma \abs{a}/b$, $\xi_{ab}/l_{Eab} = \sqrt{12}$.}
  \label{fig:phi_dependences}
\end{figure}

In Fig. \ref{fig:phi_dependences} we plot the analytical and numerical $\varphi$ dependencies of the diagonal components of the full viscosity ($\hat{\eta} = \hat{\eta}_{p0} + \hat{\eta}_{oh}$). 

\section{Temperature dependence of the viscosity anisotropy.}
\label{sec:GTDGL}

Within the framework of the TDGL equation \eqref{eq:TDGL} the viscosity anisotropy $\eta_{xx}/\eta_{yy}$ does not depend on temperature. However, the region of applicability of Eq. \eqref{eq:TDGL} is limited by gapless superconductivity. In this section we consider a more general approach based on the generalized TDGL equations\cite{Watts-Tobin+81} (see also Ref. \onlinecite{Kopnin_book} for review):
\begin{equation}
    2 \hbar \gamma \sqrt{1 + q \abs{\psi}^2/\abs{\psi_{\infty}}^2} \frac{\partial \abs{\psi}}{\partial t} = -\frac{\delta F}{\delta \abs{\psi}},
    \label{eq:amplitude}
\end{equation}
\begin{equation}
    \frac{\gamma \abs{\psi}^2}{\sqrt{1 + q \abs{\psi}^2 \! /\abs{\psi_{\infty}}^2}} \! \left( \! \hbar \frac{\partial \theta}{\partial t} + 2e \Phi \! \right) \! = \!\frac{\hbar^2}{2} \nabla \! \left( \abs{\psi}^2 \hat{m}^{-1} \nabla \theta \right) \!,
    \label{eq:phase}
\end{equation}
\[ q=\frac{32 \pi^2 \tau_{\mathrm{ph}}^2 T_c (T_c - T)}{7 \zeta(3) \hbar^2}, \qquad \abs{\psi_{\infty}}^2 = \frac{\abs{a}}{b}.\]
Here $T_c$ is the critical temperature and $\tau_{\mathrm{ph}}$ is the electron-phonon mean free time. In the isotropic case Eqs. \eqref{eq:amplitude} and \eqref{eq:phase} are valid for dirty superconductors, when the temperature is close to $T_c$ and variations of the order parameter in space and in time are sufficiently slow.

The main relations for the viscosity can be derived in same way as described in section \ref{sec:maineq}. As a result, we find that the viscosity still comprises two terms representing two mechanisms of dissipation, but the viscosity components undergo some changes. For example, Eq. \eqref{eq:eta_p0} is modified as follows:
\begin{equation}
        (\eta'_{p0})_{ij} = 2 \pi \hbar \gamma \frac{\abs{a}}{b} \delta_{ij} \int_0^{\infty} \left( \frac{df}{d \rho} \right)^2 \rho \sqrt{1 + q f^2(\rho)} d \rho.
    \label{eq:eta_p0_Gen}
\end{equation}
In order to obtain the counterparts of Eqs. \eqref{eq:eta_xx}, \eqref{eq:eta_yy}, \eqref{eq:Phi_x} and \eqref{eq:Phi_y} one should make the following substitutions in these equations:
\begin{eqnarray}
 u^2 \quad & \to &  \quad \frac{u^2}{\sqrt{1 + q}}, \nonumber \\
 f^2 \quad & \to & \quad \frac{\sqrt{1 + q} f^2}{\sqrt{1 + q f^2}}, \nonumber \\
 \eta_i \quad & \to & \quad \eta_i \sqrt{1+q}, \qquad i=x,y.
 \label{eq:substitutions}
\end{eqnarray}
It can be seen from Eq. \eqref{eq:phase} that the electric field penetration depth is increased by a factor $(1+q)^{1/4}$ as compared to Eq. \eqref{eq:xi_lE}. It may seem that at low temperatures we would reach the $l_E \gg \xi$ limit, which has been analysed in Ref. \onlinecite{Genkin+89}. However, this is not quite true because of the different relative impacts of the two mentioned mechanisms of dissipation in the simple and generalized TDGL models. Within the simple TDGL theory the Bardeen-Stephen contribution and the relaxational term are of the same order of magnitude in the $l_E \gg \xi$ limit. On the contrary, in the generalized model the viscosity is dominated by the relaxational term at low temperatures (see below).

It is obvious that all main relations from sec. \ref{sec:approximate} can be derived again within the generalized TDGL theory, but they are slightly modified. For example, Eq. \eqref{eq:eta_short} now reads
\begin{equation}
    \etaxx = \etayy \approx 2\pi \hbar \gamma \frac{\abs{a}}{b} \left[ \frac{\sqrt{k_2}}{u} - \frac{1+2q}{8u^2} \right].
    \label{eq:s1_general}
\end{equation}

\begin{figure}[t]
  \hspace*{-15pt}
  \includegraphics[scale=0.11]{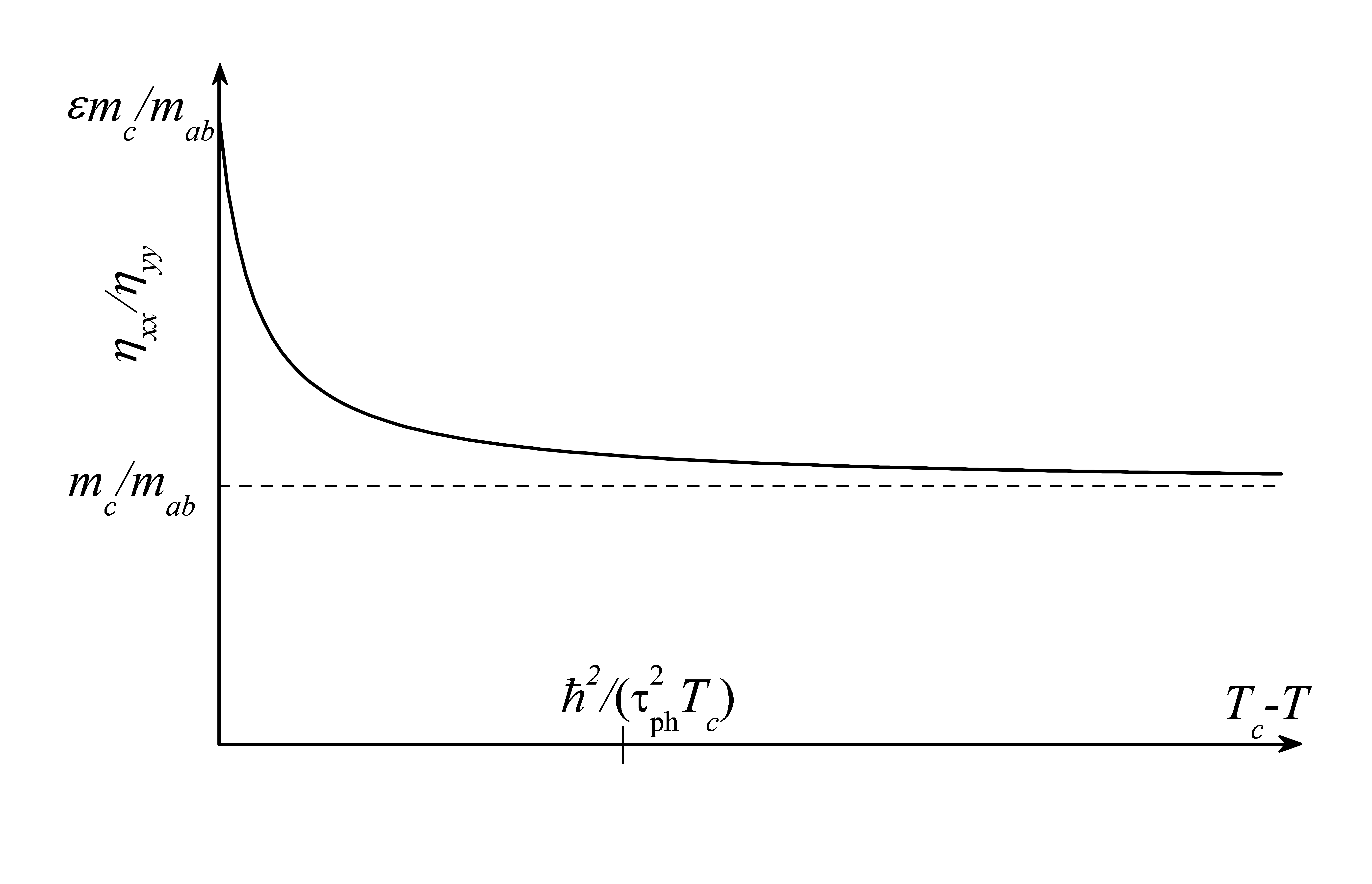}
  \caption{Schematic temperature dependence of the viscosity anisotropy for $\varphi = \pi/2$. The parameter $\epsilon$ is of the order of unity.} \label{fig:anisotropy}
\end{figure}

Now consider the temperature dependence of the viscosity. The quantity $q$ depends on the temperature $T$, and $q'_T<0$. Hence,
\[ \frac{\partial}{\partial T} \frac{(\eta_{p0})_{xx}}{\eta_0} = \frac{\partial}{\partial T} \frac{ (\eta_{p0})_{yy}}{\eta_0} < 0, \]
where $\eta_0 = \hbar \gamma \abs{a} /b$. On the other hand, the modified Eq. \eqref{eq:tilde_eta} can be written in the form
\begin{eqnarray}
    &\tilde{\eta}(s_x,s_y,u) = 2 \frac{ \left| a \right|}{b} \gamma \hbar u^2 \Min{\phi} \int \left[ s_x \left(\frac{\partial \phi}{\partial x} \right)^2 \right. & \nonumber \\
    &\left. + s_y \left(\frac{\partial \phi}{\partial y} \right)^2 + \frac{f^2(\rho)}{u^2 \sqrt{1+q f^2}} \left( u^2 \phi - \frac{y}{\rho^2} \right)^2 \right] dx \, dy,&
    \label{eq:tilde_eta_gen}
\end{eqnarray}
if we leave Eqs. \eqref{eq:eta_var'} unchanged. Hence,
\[ \frac{\partial}{\partial T} \frac{(\eta_{\mathrm{oh}})_{xx}}{\eta_0}>0; \qquad \frac{\partial}{\partial T} \frac{ (\eta_{\mathrm{oh}})_{yy}}{\eta_0} > 0. \]
At sufficiently low temperatures, when $q \gg 1$, $s \lesssim 1$ and $u \sim 1$, it may happen that
\[ (\eta_{p0})_{xx} \gg (\eta_{\mathrm{oh}})_{xx}, \qquad (\eta_{p0})_{yy} \gg (\eta_{\mathrm{oh}})_{yy}. \]
Then the viscosity anisotropy is determined by the relaxational term:
\[ \frac{\eta_{xx}}{\eta_{yy}} \approx \frac{(\eta_{p0})_{xx}}{(\eta_{p0})_{yy}}= \frac{1+\mu}{1+\mu \cos^2 \varphi}. \]
Note that when $s \neq 1$ $\eta_x \neq \eta_y$, so
\[ \frac{\eta_{xx}}{\eta_{yy}} \neq \frac{1+\mu}{1+\mu \cos^2 \varphi} \]
when $q \lesssim 1$. We have proved that within the generalized
TDGL theory the viscosity anisotropy  and the flux-flow
conductivity anisotropy do depend on temperature. The
schematic $T$ dependence of the ratio
$\eta_{xx}/\eta_{yy}$ is plotted in Fig. \ref{fig:anisotropy}.
\section{Conclusion}

By solving the time-dependent Ginzburg-Landau equation we analyzed
the viscous flux-flow in anisotropic superconductors. The
Bardeen-Stephen contribution to the viscosity tensor $\hat{\eta}$
has been calculated in the $l_{E} \ll \xi$ and $l_{Ec} \gg \xi_c$
limits. We emphasize that in these calculations we did not use any
simplifying assumptions concerning the shape of the order
parameter in a static vortex. We suggested a variational
procedure, which allowed us to derive the relations
\eqref{etaxx_universal} and \eqref{etayy_universal} suitable for
arbitrary electric field penetration lengths ($l_{Eab}$ and
$l_{Ec}$), coherence lengths ($\xi_{ab}$ and $\xi_c$) and
orientation of the magnetic field. Our results may be
useful for interpretation of experimental data on flux flow
conductivity in isotropic and anisotropic superconductors in weak
magnetic fields ($B \ll H_{c2}$).

Viscous flux-flow has also been examined within a generalized TDGL theory. We found that the viscosity anisotropy may depend on temperature and, thus, the flux-flow conductivity anisotropy may be altered by heating or cooling the sample. We hope that this effect will be observed experimentally in the near future.

\section{Acknowledgements}

This work was supported, in part, by the Russian Foundation for
Basic Research,
 Russian Agency of Education under the Federal Program ``Scientific and educational personnel of innovative Russia in 2009-2013'', and European IRSES program SIMTECH (contract n. 246937).
We are thankful to N.B. Kopnin for helpful discussions.

\appendix
\section{}
\label{app:basic}

In this appendix we derive Eqs. \eqref{eq:eta_xx}, \eqref{eq:eta_yy}, \eqref{eq:Phi_x} and \eqref{eq:Phi_y}.
In Eq. \eqref{eq:TDGL} it is convenient to make a scaling of the variables: $x'=x  (m(\varphi)/m_{ab})^{1/2}$, $y' = y$, $z'=z$. We rewrite Eq. \eqref{eq:TDGL} in the form
\begin{equation}
    \gamma \hbar \frac{\partial \abs{\psi}}{\partial t} = \frac{\hbar^2}{2m_{ab}} \left[ \nabla'^2 \abs{\psi} - \abs{\psi} (\nabla' \theta)^2 \right] - a \abs{\psi} - b \abs{\psi}^3,
    \label{eq:TDGL_re}
\end{equation}
\begin{equation}
    \gamma \abs{\psi}^2 \left( \hbar \frac{\partial \theta}{\partial t} + 2e \Phi \right) = \frac{\hbar^2}{2m_{ab}}\nabla' ( \abs{\psi}^2 \nabla' \theta).
    \label{eq:TDGL_im}
\end{equation}
According to Eq. \eqref{eq:divj}, the two-dimensional current $\mathbf{j}'~= (j_x [m(\varphi)/m_{ab}]^{1/2}, j_y)$ satisfies the relation
\begin{equation}
    \mathrm{div}'\mathbf{j}' = 0.
    \label{eq:divj'}
\end{equation}
It follows from Eqs. \eqref{eq:TDGL_im} and \eqref{eq:divj'} that
\begin{equation}
    \gamma \abs{\psi}^2 \left( \hbar \frac{\partial \theta}{\partial t} + 2e \Phi \right) = \frac{\hbar}{4e} \nabla' \left( \tilde{\sigma}_{n} \nabla' \Phi \right),
    \label{eq:Phi_main}
\end{equation}%
where we introduced the tensor $\tilde{\sigma}_{n}$ with components
\begin{eqnarray}
    &\tilde{\sigma}_{nx'x'} = s(\varphi) \sigma_{ab}, \quad \tilde{\sigma}_{ny'y'}=\sigma_{ab}, &\nonumber \\
    & \tilde{\sigma}_{nx'y'} = \tilde{\sigma}_{ny'x'} = 0, &
    \label{eq:sigma}
\end{eqnarray}
with $s(\varphi)$ given by Eq. \eqref{eq:s}. For a moving vortex one should search the solution of Eqs.
\eqref{eq:TDGL_re}, \eqref{eq:TDGL_im} and \eqref{eq:divj'} in the
form $\psi = \psi(\boldsymbol{\rho'} - \tildeVL t)$, $\Phi =
\Phi(\boldsymbol{\rho'} - \tildeVL t)$, where $\mathbf{\tilde{V}_L} =
(V_{Lx} [m(\varphi)/m_{ab}]^{1/2},~V_{Ly})$ and $\boldsymbol{\rho'}
= (x', y')$.
We expand $\abs{\psi}$ and $\theta$ in powers of $V_L$ up to the first order term, assuming the vortex velocity to be sufficiently small:
\begin{equation}
    \abs{\psi} \approx \psi_0(\boldsymbol{\rho'} - \tildeVL t) + \psi_1(\boldsymbol{\rho'} - \tildeVL t),
    \label{eq:psi_1}
\end{equation}
\begin{equation}
    \theta \approx \theta_0(\boldsymbol{\rho'} - \tildeVL t) + \theta_1(\boldsymbol{\rho'} - \tildeVL t).
    \label{eq:theta1}
\end{equation}
Here $\psi_0(\boldsymbol{\rho'})$ and $\theta_0(\boldsymbol{\rho'})$ correspond to a static vortex. The functions $\psi_1$, $\theta_1$ and $\Phi$ are of the order $V_L$. We substitute \eqref{eq:psi_1} and \eqref{eq:theta1} into Eqs. \eqref{eq:TDGL_re}, \eqref{eq:divj'}  and \eqref{eq:Phi_main}:
\begin{equation}
    -a \psi_0 - b \psi_0^3 + \frac{\hbar^2}{2m_{ab}} \left[ \nabla'^2 \psi_0 - \psi_0 (\nabla' \theta_0)^2 \right] = 0,
    \label{eq:psi0}
\end{equation}
\begin{eqnarray}
&\frac{\hbar^2}{2m_{ab}} \left[ \nabla'^2 \psi_1 - \psi_1 (\nabla' \theta_0)^2 - 2 \psi_0 \nabla' \theta_0 \cdot \nabla' \theta_1 \right] & \nonumber \\
    &-a \psi_1 - 3b \psi_0^2 \psi_1 = -\gamma \hbar \psiL &
    \label{eq:psi1}
\end{eqnarray}
\begin{equation}
    \mathrm{div}' \mathbf{j}'_0 =0, \qquad \mathbf{j}'_0 = \frac{2e \hbar \psi_0^2}{m_{ab}} \nabla' \theta_0,
    \label{eq:j0'}
\end{equation}
\begin{eqnarray}
    &\mathrm{div}' \mathbf{j}'_1 =0,& \nonumber \\
    & \mathbf{j}'_1 = \frac{2e \hbar}{m_{ab}} \left( 2 \psi_0 \psi_1 \nabla' \theta_0 + \psi_0^2 \nabla' \theta_1 \right) - \tilde{\sigma}_n \nabla' \Phi,&
    \label{eq:j1'}
\end{eqnarray}
\begin{equation}
    \frac{\hbar \sigma_{ab}}{4e} \left[ s(\varphi) \frac{\partial^2 \Phi}{\partial x'^2} + \frac{\partial^2 \Phi}{\partial y'^2} \right] = \gamma \psi_0^2 \left( 2e\Phi - \hbar \mathbf{\tilde{V}_L} \cdot
    \nabla' \theta_0 \right).
    \label{eq:Phi_initial}
\end{equation}
Now we introduce some new notations: $\psi_d = (\mathbf{d} \nabla') \psi_0$, $\theta_d = (\mathbf{d} \nabla') \theta_0$, $\vecjd = (\mathbf{d} \nabla') \mathbf{j}_0'$, where $\mathbf{d}$ is an arbitrary vector. A simple equation connecting $\psi_d$ and $\theta_d$ can be obtained by applying the operator $\mathbf{d} \nabla'$ to Eq. \eqref{eq:psi0}:
\begin{eqnarray}
&\frac{\hbar^2}{2m_{ab}} \left[ \nabla'^2 \psi_d - \psi_d (\nabla' \theta_0)^2 - 2 \psi_0 \nabla' \theta_0 \cdot \nabla' \theta_d \right] & \nonumber \\
    &-a \psi_d - 3b \psi_0^2 \psi_d = 0 &
    \label{eq:psi_d}
\end{eqnarray}
The vector $\vecjd$ satisfies the obvious relation $\mathrm{div}' \vecjd =0$. Let us multiply Eq. \eqref{eq:psi1} by $\psi_d$, subtract Eq. \eqref{eq:psi_d} multiplied by $\psi_1$ and integrate the resulting equation over a large volume containing the whole vortex. After some simple algebra and integration by parts we obtain
\begin{eqnarray}
& -\gamma \hbar \int \psiL \psi_d d^3 \mathbf{r}' \qquad &\nonumber\\
 & \qquad = \frac{\hbar}{4e} \int \left[ (\mathbf{j}'_1 + \tilde{\sigma}_n \nabla' \Phi) \nabla' \theta_d - \vecjd \nabla' \theta_1 \right] d^3 \mathbf{r}' & \nonumber\\
 & = \frac{\hbar}{4e} \int (\tilde{\sigma}_n \nabla' \Phi) \nabla' \theta_d  d^3 \mathbf{r}' + \frac{\hbar}{4e} \int_S (\mathbf{j}'_1 \theta_d - \vecjd \theta_1) d \mathbf{S}. \qquad &
\label{eq:unevaluated}
\end{eqnarray}
Here $S$ is a surface far from the vortex axis. At large distances $\rho' \gg \xi_{ab}$ we have
%
\[ \mathbf{j}'_1 \approx \frac{2e \hbar \abs{a}}{b m_{ab}} \nabla' \theta_1 = \mathbf{j}'_{\mathrm{tr}}, \quad \theta_1 = \frac{b m_{ab}}{2e \hbar \abs{a}} (\mathbf{j}'_{\mathrm{tr}} \cdot \boldsymbol{\rho'}) + \mathrm{const}, \]
where $\mathbf{j}'_{\mathrm{tr}}$ is the transport current which is constant. If we calculate the surface integral in Eq. \eqref{eq:unevaluated} and make some simple transformations, we obtain the force balance equation\cite{Genkin+89}
\begin{eqnarray}
    &\frac{\pi \hbar}{e} \left[ \mathbf{d} \cdot ( \mathbf{j}'_{\mathrm{tr}} \times \mathbf{n} ) \right] = - 2 \pi \gamma \hbar (\mathbf{d} \cdot \tildeVL)  \int_0^{\infty} \left( \frac{d \psi_0}{d\rho} \right)^2 \rho \, d\rho & \nonumber \\
    &+ \frac{\hbar \sigma_{ab}}{2e} \int \left[  s(\varphi) \frac{\partial^2 \Phi}{\partial x'^2} + \frac{\partial^2 \Phi}{\partial y'^2} \right] (\vecd \cdot \nabla' \theta_0) d^2 \boldsymbol{\rho}',&
    \label{eq:evaluated}
\end{eqnarray}
where  $\mathbf{n}$ is a unit vector along the magnetic field.
If we compare Eqs. \eqref{eq:vortex_motion} and \eqref{eq:evaluated}, we can see that the viscosity tensor in the frame $(x',y',z')$ should be defined as follows:
\begin{eqnarray}
    \lefteqn{\mathbf{d} \cdot \hat{\eta}' \tildeVL = 2 \pi \gamma \hbar (\mathbf{d} \cdot \tildeVL) \frac{\abs{a}}{b} \int_0^{\infty} \left( \frac{df}{d\rho} \right)^2 \rho d\rho} & & \nonumber \\
    & &- \frac{\hbar \sigma_{ab}}{2e} \int \left[  s(\varphi) \frac{\partial^2 \Phi}{\partial x'^2} + \frac{\partial^2 \Phi}{\partial y'^2} \right] (\vecd \cdot \nabla' \theta_0) d^2 \boldsymbol{\rho}',
    \label{eq:eta_summ}
\end{eqnarray}
where we introduced the function

\begin{equation}
	f(\rho) = \psi_0 (\rho \xi_{ab}) \sqrt{\frac{b}{\abs{a}}}.
	\label{eq:f_def}
\end{equation}
The components of the viscosity $\hat{\eta}$ in the frame $(x, \, y, \, z)$ are given by
\begin{equation}
 \eta_{xx} = [m(\varphi)/m_{ab}]^{1/2} \eta'_{x'x'}, \quad \eta_{yy} = [m_{ab}/m(\varphi)]^{1/2} \eta'_{y'y'}
    \label{eq:eta_transform}
\end{equation}
The rhs of Eq. \eqref{eq:eta_summ} contains two terms, representing two mechanisms of dissipation. The viscosity due to relaxation of the order parameter is\cite{Hu+72,Hu72,Kupriyanov+72,Genkin+89}:
\begin{equation}
    (\eta'_{p0})_{ij} = 2 \pi \hbar \gamma \frac{\abs{a}}{b} \alpha_1 \delta_{ij},
    \label{eq:eta_p0}
\end{equation}
\[ \alpha_1 = \int_0^{\infty} \left( \frac{df}{d\rho} \right)^2 \rho d\rho = 0.279. \]
The second term in the rhs of Eq. \eqref{eq:eta_summ} defines the ohmic viscosity tensor $\hat{\eta}'_{oh}$, which is to be evaluated:
\begin{equation}
    \mathbf{d} \cdot \hat{\eta}'_{oh} \tildeVL = - \frac{\hbar \sigma_{ab}}{2e} \int \left[  s(\varphi) \frac{\partial^2 \Phi}{\partial x'^2} + \frac{\partial^2 \Phi}{\partial y'^2} \right] (\vecd \nabla' \theta_0) d^2 \boldsymbol{\rho}'.
    \label{eq:ohmic}
\end{equation}
Now, if we substitute $\Phi$ in the form \eqref{eq:PhixPhiy} into Eqs. \eqref{eq:Phi_initial} and \eqref{eq:ohmic} and switch to the coordinates $(x_1,y_1)$ (see Eq. \eqref{eq:x1y1}) we obtain Eqs. \eqref{eq:eta_xx}, \eqref{eq:eta_yy}, \eqref{eq:Phi_x} and \eqref{eq:Phi_y}.
\section{}
\label{app:largeu}

In this Appendix we will derive Eq. \eqref{eq:etaxx_final}. First, we divide the integral in Eq. \eqref{eq:eta_xx} into two parts:
\begin{equation}
    \etaxx = \eta_{x1} + \eta_{x2},
    \label{eq:etaxx_summ}
\end{equation}
\begin{equation}
     \eta_{x1} = -2 \frac{ \left| a \right|}{b} \gamma \hbar \!\! \int_{\rho_1 < \rho_0 / \sqrt{u}}  \!\!\!\!\!\!\!\! {f^2(\rho_1) \frac{y_1}{\rho_1^2} \left( u^2 \Phi_x - \frac{y_1}{\rho_1^2} \right) \! dx_1 dy_1},
     \label{eq:etaxx1}
\end{equation}
\begin{equation}
    \eta_{x2} = -2 \frac{ \left| a \right|}{b} \gamma \hbar \!\! \int_{\rho_1 > \rho_0 / \sqrt{u}} \!\!\!\!\!\!\!\! { f^2(\rho_1) \frac{y_1}{\rho_1^2} \left( u^2 \Phi_x - \frac{y_1}{\rho_1^2} \right) \! dx_1 dy_1,}
    \label{eq:etaxx2}
\end{equation}
where $\rho_0 = u^{1/6+\delta}$, $\delta \ll 1/6$. Note that the lhs of Eq. \eqref{eq:Phi_x} is small when $\rho_1 > \rho_0 / \sqrt{u} \gg u^{-1/2}$, so it can be accounted for by perturbation theory:
\begin{equation}
    \Phi_x = \frac{y_1}{u^2 \rho_1^2} + \Phi_{x1} + \Phi_{x2} + ...,
    \label{eq:Phix_expand}
\end{equation}
\[ \Phi_{x1} = \frac{1}{u^4 f^2(\rho_1)} \left( \frac{\partial^2}{\partial y_1^2} + s \frac{\partial^2}{\partial x_1^2} \right) \frac{y_1}{\rho_1^2}, \]
\[ \Phi_{x2} = \frac{1}{u^6} \left[ \frac{1}{f^2(\rho_1)} \left( \frac{\partial^2}{\partial y_1^2} + s \frac{\partial^2}{\partial x_1^2} \right) \right]^2  \frac{y_1}{\rho_1^2}. \]
The main contribution to the integral in Eq. \eqref{eq:etaxx2} is determined by small $\rho_1$. The integral of $\Phi_{x2}$ is of the order of $(u \rho_0^6)^{-1} \ll u^{-2}$, the integrals of higher-order terms are also negligibly small, hence
\begin{equation}
    \eta_{x2} \approx -2 \frac{\abs{a}}{b} \gamma \hbar \frac{I_{0x}'}{u} ,
    \label{eq:etaxx2_final}
\end{equation}
\[ I_{0x}' = \int_{\rho > \rho_0}  \frac{y}{\rho^2} \left( \frac{\partial^2}{\partial y^2} + s \frac{\partial^2}{\partial x^2} \right) \frac{y}{\rho^2} \,dx \, dy. \]
Let us consider the component $\eta_{x1}$. In the new variables introduced in subsection \ref{sub:large_u} Eq. \eqref{eq:etaxx1} reads
\begin{equation}
    \eta_{x1} = -2 \frac{ \left| a \right|}{b} \gamma \hbar \int_{\rho < \rho_0}  \!\!\! {f^2 \left( \frac{\rho}{\sqrt{u}} \right) \frac{y}{\rho^2} \left( \tilde{\Phi}_x - \frac{y}{\rho^2} \right) \, dx \, dy}.
     \label{eq:etaxx1_1}
\end{equation}
Now we estimate the term $R_x$ introduced in Eq. \eqref{eq:Phix_expand1}. It satisfies the following relation:
\begin{eqnarray} 
 &\frac{\partial^2 R_x}{\partial y^2} + s \frac{\partial^2 R_x}{\partial x^2}- u f^2 \left( \frac{\rho}{\sqrt{u}} \right) R_x & \nonumber \\  
	&= \left[ u f^2\left( \frac{\rho}{\sqrt{u}} \right) - k_2 \rho^2 - \frac{k_4 \rho^4}{u} \right] \left( \Phixnull - \frac{y}{\rho^2} \right) & \nonumber \\
	&+ \left[ u f^2\left( \frac{\rho}{\sqrt{u}} \right) - k_2 \rho^2 \right] \frac{\Phixone}{u}.&
	\label{eq:R_x}
\end{eqnarray}
Note that when $\rho \ll \sqrt{u}$ the source in the rhs of \eqref{eq:R_x} can be presented as $S(x,y)u^{-2}$, where $S(x,y)$ is some function independent of $u$. Since \eqref{eq:R_x} is a screening equation, the function $R_x(x,y,u)$ for small $\rho$ does not depend on the behavior of the source in the area of big $\rho$ and can be presented as $R_x = \tilde{R}_x(x,y)u^{-2}$. On the other hand, when $\rho \gg 1$ the derivatives in the lhs of Eq. \eqref{eq:R_x} are small, hence in the area $1 \ll \rho \ll \sqrt{u}$
\[ R_x \approx \frac{1}{u^2} \left[ -\frac{k_6 \rho^4}{k_2} \left( \Phixnull - \frac{y}{\rho^2} \right) - \frac{k_4 \rho^2}{k_2} \Phixone \right], \]
\begin{equation}
	\abs{R_x} \leq \frac{\mathrm{const}}{\rho u^2}.
	\label{eq:R_x_estim}
\end{equation}
Now we substitute $\tilde{\Phi}_x$ in the form \eqref{eq:Phix_expand1} into \eqref{eq:etaxx1_1}:

\begin{equation}
	\eta_{x1} = -2 \frac{\abs{a}}{b} \gamma \hbar \left( \frac{I_{1x}'}{u}   + \frac{I_{2x}'}{u^2} + I_{3x} \right),
	\label{eq:etaxx1_final}
\end{equation}
where 
\begin{equation}
    I_{1x}' = \int_{\rho < \rho_0} k_2 y \left( \Phixnull - \frac{y}{\rho^2} \right) dx \, dy,
    \label{eq:I_1x'}
\end{equation}
\begin{equation}
        I_{2x}' = \!\! \int_{\rho < \rho_0} \! \frac{y}{\rho^2} \left[ k_4 \rho^4 \left( \Phixnull - \frac{y}{\rho^2} \right) \! + k_2 \rho^2 \Phixone \right]  dx \, dy,
        \label{eq:I_2x'}
\end{equation}
\begin{eqnarray}
	 &I_{3x} = \! \int_{\rho < \rho_0} \! \left[ \fwithu - \frac{k_2 \rho^2}{u} - \frac{k_4 \rho^4}{u^2} \right] \!\! \left( \Phixnull \! - \frac{y}{\rho^2} \right) \frac{y}{\rho^2} dx \, dy & \nonumber \\
 &+ \int_{\rho < \rho_0} \left[ \fwithu - \frac{k_2 \rho^2}{u} \right] \frac{\Phixone}{u} \frac{y}{\rho^2} dx \, dy & \nonumber\\
 &+ \int_{\rho < \rho_0} \fwithu R_x \frac{y}{\rho^2} dx \, dy.&
\label{eq:I_3x}
\end{eqnarray}
One can easily prove that
\begin{equation}
    \Phixnull = \frac{y}{\rho^2} + \frac{1}{k_2 \rho^2} \left( \frac{\partial^2}{\partial y^2} + s \frac{\partial^2}{\partial x^2} \right) \frac{y}{\rho^2} + O( \rho^{-9}),
    \label{eq:Phixnull_asymp}
\end{equation}
\begin{equation}
     k_4 \rho^4 \left( \Phixnull - \frac{y}{\rho^2} \right) + k_2 \rho^2 \Phixone = O( \rho^{-5}).
     \label{eq:Phixone_asymp}
\end{equation}
From Eqs. \eqref{eq:Phixnull_asymp}, \eqref{eq:Phixone_asymp} and \eqref{eq:R_x_estim} we can see that all integrals in Eq. \eqref{eq:I_3x} are of the order $\rho_0^2 / u^3$. Thus $\abs{I_{3x}} \ll u^{-2}$, so it can be neglected.
Also we can integrate in Eqs. \eqref{eq:I_1x'} and \eqref{eq:I_2x'} over the whole $xy$ plane, since
\begin{equation}
     \abs{I'_{0x} + I'_{1x} - I_{1x}(s)} \ll u^{-1},
     \label{eq:I_1x}
\end{equation}
\begin{equation}
    \abs{I'_{2x} - I_{2x}(s)} \ll 1,
    \label{eq:I_2x}
\end{equation}
Finally, taking into account Eqs. \eqref{eq:etaxx_summ}, \eqref{eq:etaxx2_final} and \eqref{eq:etaxx1_final} we obtain Eq. \eqref{eq:etaxx_final}.
\section{}
\label{app:sggu2}

In this Appendix we consider in detail the derivation of Eqs.
\eqref{eq:eta_x_s_final} and \eqref{eq:eta_y_s_final}. We will
present here the calculations for the $\etayy$ component, since
the calculations for the $\etaxx$ component are less complicated.
First, we rewrite Eq. \eqref{eq:Phi_y} in the form
\begin{eqnarray}
    &s \frac{\partial^2 \Phi_y}{\partial x^2} - u^2 \Phi_y + \frac{x}{\rho^2} f^2(\rho)  = \qquad \qquad & \nonumber \\
    & \qquad \qquad =-\frac{\partial^2 \Phi_y}{\partial y^2} - u^2 \Phi_y \left( 1 - f^2(\rho) \right).&
    \label{eq:Phi_y_s}
\end{eqnarray}
The index ``1'' is omitted. It will be proved below that the terms in the rhs of Eq. \eqref{eq:Phi_y_s} give a small contribution to the viscosity, so they can be neglected. Then the solution of Eq. \eqref{eq:Phi_y_s} has the form
\begin{equation}
    \Phi_y \approx \Phi_{y0} =  \int_{-\infty}^{+\infty} \frac{x'f^2(\rho')}{x'^2 + y^2}  \frac{\exponent}{2u \sqrt{s}} dx',
    \label{eq:Phi_y_solution}
\end{equation}
where $\rho'= \sqrt{x'^2 + y^2}$. Consider a quantity $y_0$ in the range $1 \ll y_0 \ll \sqrt{s}/u$ (for example, $y_0 = s^{1/4}/u^{1/2}$). We divide the integral in Eq. \eqref{eq:eta_yy} into three parts:
\begin{eqnarray}
    &\etayy = -2 \frac{ \left| a \right|}{b} \gamma \hbar \left[
    \int_{\left| y \right| < y_0}{ f^2(\rho) \frac{x}{\rho^2} u^2 \Phi_y \, dx \, dy }  - \right. &\nonumber \\
    &\left. -\int_{\left| y \right| < y_0}{ f^2(\rho) \frac{x^2}{\rho^4} \, dx \, dy } \right] &\nonumber \\
    &- 4 \frac{ \left| a \right|}{b} \gamma \hbar \int_{y > y_0}{ f^2(\rho) \frac{x}{\rho^2} \left( u^2 \Phi_y - \frac{x}{\rho^2} \right) \, dx \, dy }.&
    \label{eq:eta_y_sum}
\end{eqnarray}
Using the inequality
\begin{equation}
    f^2(\rho) < \frac{\rho^2}{A_1 + \rho^2},
    \label{eq:A_1}
\end{equation}
where $A_1$ is some constant, we can estimate the first integral:
\begin{equation}
	\abs{\int_{\left| y \right| < y_0}{ f^2(\rho) \frac{x}{\rho^2} u^2 \Phi_y \, dx dy }} 
\leq \const \frac{y_0 u}{\sqrt{s}} \left( \ln{\frac{\sqrt{s}}{u}} \right)^2 \ll 1
\end{equation}
Here and further ``$\const$'' denotes a constant independent of any parameters. The second term in Eq. \eqref{eq:eta_y_sum} has the following asymptotics when $y_0 \gg 1$:
\begin{equation}
     \int_{\left| y \right| < y_0}{ f^2(\rho) \frac{x^2}{\rho^4} \, dx \, dy } \approx \pi \ln y_0 + C_y.
     \label{eq:2term}
\end{equation}
The constant $C_y$ will be evaluated below. The third integral in
Eq. \eqref{eq:eta_y_sum} can be simplified if we take into account
that $\rho>y_0$, $\rho'>y_0$ and $y_0 \gg 1$, so we can substitute
unity instead of $f^2$:
\begin{eqnarray}
& \int_{y > y_0} f^2(\rho) \frac{x}{\rho^2} \left( u^2 \Phi_y - \frac{x}{\rho^2} \right) \, dx \, dy & \nonumber \\
& \approx \int_{y_0}^{\infty} \left( \frac{\pi u}{\sqrt{s}} \inftyint \frac{y \exp\left(-\frac{u\abs{x}}{\sqrt{s}} \right)}{x^2 + 4y^2} dx - \frac{\pi}{2y} \right) \! dy \! & \nonumber \\
& =  \frac{\pi}{2} \int_0^{\infty} dx \int_{y_0}^{\infty} dy \left( \frac{4y}{sx^2/u^2 +4y^2} - \frac{1}{y} \right) e^{-x} & \nonumber \\
& \approx -\frac{\pi}{4} \int_0^{\infty} \ln \frac{sx^2}{4y_0^2 u^2} e^{-x} dx 
 =\frac{\pi}{2} \ln{y_0} - \frac{\pi}{4} \ln{\frac{s}{4u^2}} + \frac{\pi}{2} {\cal C}, \qquad &
     \label{eq:3term}
\end{eqnarray}
where $\cal C$ is the Euler constant:
\[ {\cal C} = - \int_0^{\infty} \ln{x} \cdot e^{-x} dx \approx 0.577. \]
Using \eqref{eq:eta_y_sum} - \eqref{eq:3term} we obtain
\begin{equation}
        \etayy = 2\pi \hbar \gamma \frac{\left| a \right|}{b} \left( \frac{1}{2} \ln \frac{s}{4u^2} + \frac{C_y}{\pi} - {\cal C} \right).
    \label{eq:eta_y_s}
\end{equation}
The component $\etaxx$ can be calculated in a similar way:
\begin{equation}
    \etaxx = 2\pi \hbar \gamma \frac{\left| a \right|}{b} \left( \frac{1}{2} \ln \frac{s}{4u^2} + \frac{C_x}{\pi} - {\cal C} \right),
    \label{eq:eta_x_s}
\end{equation}
\begin{equation}
    C_x = \lim_{y_0 \rightarrow \infty} \left( \int_{\left| y \right| < y_0}{ f^2(\rho) \frac{y^2}{\rho^4} \, dx dy } - \pi \ln y_0 \right).
    \label{eq:C_x}
\end{equation}

Now we evaluate $C_x$ and $C_y$. Here the constant $g'_4$ from Ref. \onlinecite{Hu72} will be useful:
\begin{equation}
    g'_4 = \int_0^{\infty} \left[ f^2(\rho) - \frac{\rho^2}{1+\rho^2} \right] \rho^{-1} d\rho =  -0.3982.
    \label{eq:g'_4}
\end{equation}
It is easy to check that
\[ C_x = \pi g'_4 + \lim_{y_0 \rightarrow \infty} \! \int_{\left| y \right| < y_0, \rho>y_0} \!\!\!\!\!\!\!\!\!\!\!\!\!\! y^2\rho^{-4} \, dx \, dy = \pi \left( g'_4 + \ln 2 - \frac{1}{2} \right) \! . \]
Similarly,
\[ C_y = \pi \left( g'_4 + \ln 2 + \frac{1}{2} \right). \]
Finally, the components of the viscosity take the form
\begin{equation}
    \etaxx = 2\pi \hbar \gamma \frac{\left| a \right|}{b} \left( \ln \frac{\sqrt{s}}{u} + g'_4 - {\cal C} -\frac{1}{2} \right),
    \label{eq:eta_x_s_final'}
\end{equation}
\begin{eqnarray}
    \etayy = 2\pi \hbar \gamma \frac{\left| a \right|}{b} \left( \ln \frac{\sqrt{s}}{u} + g'_4 - {\cal C} +\frac{1}{2} \right).
    \label{eq:eta_y_s_final'}
\end{eqnarray}
If we substitute ${\cal C}$ and $g'_4$ with their numerical values, we obtain Eqs. \eqref{eq:eta_x_s_final} and \eqref{eq:eta_y_s_final}.

Now it is necessary to prove our assumption concerning the
rhs of Eq. \eqref{eq:Phi_y_s}. Consider it as a
perturbation. The first order correction to the approximate solution $\Phi_{y0}$ has the form 
\[ \Phi_{y1} = R_{y}^{(1)} + R_{y}^{(2)},\]
\begin{equation}
     R_{y}^{(1)} = \!\! \inftyint  \! \frac{u \Phi_{y0}(x',y) [1 - f^2(\rho')]}{2 \sqrt{s}} \exp \!\! \left( \!\! -\frac{u\abs{x-x'}}{\sqrt{s}} \right) dx',
    \label{eq:Ry1}
\end{equation}
\begin{equation}
    R_{y}^{(2)} = \frac{1}{2u \sqrt{s}} \inftyint \frac{\partial^2 \Phi_{y0}}{\partial y^2}(x',y)  \exp \left( -\frac{u\abs{x-x'}}{\sqrt{s}} \right) dx'.
    \label{eq:Ry2}
\end{equation}
The contribution of $\Phi_{y1}$ to $\etayy$ is equal to
\[ \Delta \etayy = -2 \frac{ \left| a \right|}{b} \gamma \hbar (I_y^{(1)} + I_y^{(2)}), \]
where
\begin{equation}
    I_y^{(1)} = \int f^2(\rho) \frac{x}{\rho^2} u^2 R_y^{(1)} dx \, dy ,
    \label{eq:Iy1}
\end{equation}
\begin{equation}
    I_y^{(2)} = \int f^2(\rho) \frac{x}{\rho^2} u^2 R_y^{(2)} dx \, dy.
    \label{eq:Iy2}
\end{equation}
We will show that $\abs{I_y^{(1)}} \ll 1$ and $\abs{I_y^{(2)}} \ll 1$ when $s \gg u^2$.

A simple estimate for $\abs{\Phi_{y0}}$ can be obtained with the help of \eqref{eq:A_1}:
\begin{equation}
     \abs{\Phi_{y0}} \leq \const \frac{\lnsu}{u \sqrt{s}}.
     \label{eq:Phiy0_estim}
\end{equation}
Using the inequality
\[ \frac{\rho^2}{A_2 + \rho^2} < f^2(\rho) \]
and Eq. \eqref{eq:Phiy0_estim} we can estimate $I_y^{(1)}$:
\[ \abs{I_y^{(1)}} \leq 
\const \frac{u}{\sqrt{s}} \lnsu  \ll 1. \]
For all $x'$ and $y$ we can write
\[ \abs{\frac{\partial^2}{\partial y^2} \left( \frac{1}{x'^2 + y^2} f^2(\rho') \right)} \leq \frac{\const}{(A_3 + x'^2 +y^2)^2}, \]
whence
\begin{eqnarray*}
&\abs{I_y^{(2)}} \leq
\frac{\const}{s} \int  \frac{\abs{x} \exp \left( -\frac{u\abs{x-x'}}{\sqrt{s}} \right) \exp \left( -\frac{u\abs{x''-x'}}{\sqrt{s}} \right)}{(A_3 + x''^2)(A_1 + x^2)} dx' dx'' dx & \\
& \leq \frac{\const}{s} \lnsu \int \frac{\exp \left( -\frac{u\abs{x''-x'}}{\sqrt{s}} \right)}{x''^2+ A_3} dx' dx'' \leq \frac{\const}{\sqrt{s} u} \lnsu \ll 1 &
\end{eqnarray*}
when $u \gtrsim 1$.

\end{document}